\documentclass[aps,prl,twocolumn,showpacs,groupedaddress]{revtex4}
\usepackage{graphicx}
\usepackage{dcolumn}
\usepackage{bm}
\usepackage{ae}
\usepackage{aecompl}
\usepackage{amsmath}
\usepackage{amssymb}
\usepackage{epsfig}
\usepackage{multirow}
\begin{document}

\title{Out-of-equilibrium processes in suspensions of oppositely charged colloids: liquid-to-crystal nucleation and gel formation}


\author{Eduardo Sanz$^{1,3}$\footnote{Electronic address: {\tt
esanz@ph.ed.ac.uk}}, Chantal Valeriani$^{1,3}$, Teun Vissers$^{1}$, Andrea Fortini$^{1}$\footnote{Present address:
Department of Physics, Yeshiva University, 500 West 185th Street, New York, NY 10033, USA}, 
Mirjam E. Leunissen$^{1}$\footnote{Present address: Center for Soft
Matter Research, Physics Department, New York University, 4 Washington
Place, New York, NY 10003, USA}, Alfons van Blaaderen$^{1}$, Daan Frenkel$^{2,4}$ and Marjolein Dijkstra$^{1}$\footnote{Electronic address: {\tt
M.Dijkstra1@uu.nl}}}
\affiliation{$^{1}$Soft Condensed Matter, Debye Institute for Nanomaterials Science, Utrecht University, Princetonplein 5, 3584 CC Utrecht, The Netherlands\\
$^{2}$ FOM Institute for Atomic and Molecular Physics, Kruislaan 407, 1098 SJ Amsterdam, The Netherlands\\ 
$^{3}$ SUPA, School of Physics, 
University of Edinburgh,
JCMB King's Buildings, Mayfield Road,
Edinburgh EH9 3JZ, UK\\
$^{4}$ Department of Chemistry, University of Cambridge,  Lensfield Road, CB2 1EW, 
Cambridge,UK}


\begin{abstract}

We study the kinetics of the liquid-to-crystal transformation and of gel formation in colloidal suspensions of
oppositely charged particles. 
We analyse, by means of both computer simulations and experiments, the evolution of 
a fluid quenched  to a state point of the phase diagram where  the most stable state is either a homogeneous crystalline solid or a solid phase in contact with a dilute
gas. 
On the one hand, at high temperatures and high packing fractions, close 
to an ordered-solid/disordered-solid coexistence line,   
we find that the fluid-to-crystal pathway
does not follow the minimum free energy route. 
On the other hand, a quench to a state point far from the ordered-crystal/disordered-crystal coexistence border
is followed by a fluid-to-solid transition through the minimum free energy pathway. 
At low temperatures and  packing fractions
we observe 
that the system undergoes a gas-liquid spinodal decomposition that, at some point, arrests 
giving rise to a gel-like structure. Both our simulations and experiments suggest that increasing the interaction range 
favors crystallization over vitrification in gel-like structures.  

\end{abstract}

\maketitle


Colloidal suspensions are excellent model systems to study condensed matter physics. 
The reason is that colloids can be observed directly with current experimental 
techniques. In addition, colloids can form  structures, such as
colloidal crystals \cite{N_2005_437_7056} or gels \cite{N_2008_453_06931}, that have potential applications, for instance, in photonics or
in cosmetics. It is therefore essential to gain a better understanding in the physics governing the
spontaneous transformation of a disordered fluid phase into a self-assembled colloidal structure.
Here, we study how oppositely charged colloids assemble to form either crystals or gels. 
In particular, we are interested in the mechanism by which the fluid transforms into either phase. 
In our work, we make
use of computer simulations to study the fluid-to-crystal transition, whereas the gel formation has been studied both
numerically and experimentally. 

Quite recently, it has been demonstrated experimentally that a suspension of oppositely 
charged particles can transform into different crystal phases \cite{N_2005_437_7056}, and that 
the equilibrium phase diagram of a simple model potential for oppositely charged 
particles exhibits the same crystal structures found in experiments \cite{PRL_2006_96_018303}. 
More recently, the gas-liquid coexistence line was calculated for the aforementioned model 
potential \cite{JCP_2006_125_094502}. Taking advantage of the accurate knowledge of the equilibrium phase diagram
we study the kinetics of both fluid-solid and fluid-gel transitions in oppositely charged
colloidal suspensions.

The fluid-to-solid transition is an intriguing problem widely studied in colloid physics \cite{Nature_2001_409_1020,S_2001_292_258}. 
The reason is that,
by understanding the initial steps of crystal growth, one can aim to control the size and the structure of the 
crystallites eventually formed. The study of crystal nucleation in suspensions of hard colloidal particles
has already received quite some attention both experimentally \cite{PRE_1997_55_3054,S_2001_292_258} and by computer simulations 
\cite{Nature_2001_409_1020,PRL_2003_90_085702}.
Other colloidal systems for which the liquid-to-solid transition mechanism has been studied are, for instance,  
equally charged particles \cite{JCP_2005_123_174902,JPCM_2002_14_7667} or 
binary mixtures of hard spheres \cite{PRL_1998_80_877,JCP_2006_125_024508}. 
Here, we present the case of oppositely charged particles.
A special characteristic of the phase diagram of oppositely charged spheres is the coexistence between 
ordered and substitutionally disordered structures on a regular 3D lattice\cite{PRL_2000_85_003217,PRL_2006_96_018303}. 
This offers a good opportunity to study the 
role of an order-disorder phase transition in the liquid-to-solid nucleation pathway. 
We find that, in certain thermodynamic conditions, the formation of a disordered phase is kinetically favored, whereas the nucleation 
of the ordered phase entails a lower free energy. 

The formation of colloidal gels has also received great attention; not only because of the 
wide use of colloidal gels in cosmetics, drugs or food additives, but also because 
understanding gelation remains a fundamental
challenge. The most popular system regarding colloidal gelation is that of colloid-polymer mixtures \cite{JCP_1954_22_1255}.  
The presence of non-adsorbing polymers induces an effective attraction between the colloids, which is responsible for the gelation 
to happen \cite{FD_1995_101_65,L_2003_19_4493}. For the case of attractive particles, the mechanism by which a colloidal fluid transforms into a gel has been described 
as the arrest of a gas-liquid spinodal 
decomposition \cite{N_2008_453_06931,JCP_2005_122_224903,PRL_2007_99_098301}.  When a low density fluid separates into a very low density gas and a high density liquid by spinodal decomposition, 
a percolating pattern of dense liquid forms after the first steps of the spinodal phase separation. If the dynamics of the particles 
contained in the network-forming liquid is sufficiently slow, the pattern gets arrested, giving rise to a gel. 
Recently, it has been pointed out that mixtures of oppositely charged colloids
can also form gels\cite{JPCB_2006_110_13220,PRE_2007_76_011401,tobegelcharged}. Hereby, we present experimental evidence  confirming this statement. 
Moreover, we make use of 
computer simulations to gather evidence that the mechanism of gel formation for this system is
an arrested spinodal decomposition, as it is the case for purely attractive particles. 
Finally, we study, both  numerically and experimentally, the role of the interaction range 
on the interplay between vitrification and crystallization. Our results indicate that increasing the interaction range 
locally favors crystallization over vitrification.

\section{Methods and technical details}
\label{detailssimu}

\subsection{Computer Simulations}

In the present work, we study a symmetric binary mixture of 
equally-sized oppositely-charged particles. The number of particles
ranges from 686 (for the gel-formation simulations) to 8000 (for the Umbrella
Sampling simulations). 1000 particles were used for the spontaneous nucleation and Forward-Flux-Sampling
(FFS) runs and for some of 
the gel-formation states. 

In our computer simulations, the screened Coulomb
interaction between two colloids of diameter $\sigma$ and charge
$Ze$  is approximated by a Yukawa potential plus a repulsive core:
\begin{equation}
\label{potencialduro}
u(r)/k_BT=
\begin{cases}
\infty    & r<\sigma\\
\pm u^*\frac{\displaystyle e^{-\kappa(r-\sigma)}}{\displaystyle r/\sigma} & \sigma \le r < r_c\\
0 & r \ge r_c
\end{cases}
\end{equation}
where  $r$ is the distance between the centre-of-mass of the colloids, $u^*$ is the energy at contact in $k_BT$ units
and $r_c$ is the cut-off radius
(set at the distance where $u(r)/(u^*k_BT)=10^{-3}$). 
The energy in $k_BT$ when $r=\sigma$, $u^*$, is equal to $Z^2\lambda_B/((1+\frac{\kappa\sigma}{2})^2\sigma)$, where
$\lambda_B=e^2/\epsilon_sk_BT$ is the
Bjerrum length ($\epsilon_s$ is the dielectric constant of the
solvent) and $\kappa=\sqrt{8\pi \lambda_B \rho_{salt}}$ is the inverse
Debye screening length for a 1:1 electrolyte ($\rho_{salt}$ is the number density of
added salt ions).
To study the liquid-to-solid transition, we performed $NPT$ Monte Carlo (MC) simulations using 
Eq. \ref{potencialduro} as the Hamiltonian. In order to study the formation of gels, we performed Brownian 
Dynamics (BD) simulations in the $NVT$ ensemble, where we substituted the hard-core 
interaction of Eq. \ref{potencialduro} by a steep $u^{*}/r^{36}$ 
repulsion. 
Our units of energy and length are $u^*$ and $\sigma$ respectively. Consequently, the time units are $t^{*}=\sigma^2/(D_0 u^*)$. The time step we used for
the integration of the Langevin position equation \cite{allen_book} is 7$\cdot$10$^{-6}t^*$. 

\begin{figure} \includegraphics[clip,width=0.48\textwidth,angle=-0]{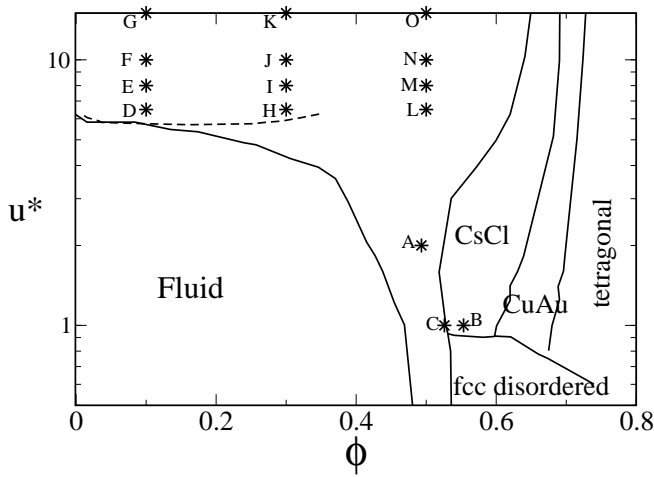}
\caption{\small Phase diagram of equally sized oppositely charged particles (taken from Refs. \cite{PRL_2006_96_018303} and \cite{JCP_2006_125_094502}). 
The Hamiltonian is given by Eq.\ref{potencialduro} ($\kappa \sigma = 6$). The solid lines indicate the coexistence lines, 
whereas the dashed line corresponds to the (metastable) gas-liquid binodal 
(solid-solid coexistence regions are too narrow to be visible on the scale of the figure). 
We have studied the structural evolution of a fluid when quenched to the state 
points indicated by an asterisk accompanied by a letter. The state points corresponding to each letter are the following:
A:( $u^* = 2$, $\phi = 0.493$), B:($u^* = 1$, $\phi = 0.553$), C:($u^*= 1 $, $\phi = 0.526$), 
D:($u^*= 6.5 $, $\phi = 0.1$),
E:($u^*= 8 $, $\phi = 0.1$),
F:($u^*= 10 $, $\phi = 0.1$),
G:($u^*= 15 $, $\phi = 0.1$),
H:($u^*= 6.5 $, $\phi = 0.3$),
I:($u^*= 8 $, $\phi = 0.3$),
J:($u^*= 10 $, $\phi = 0.3$),
K:($u^*= 15 $, $\phi = 0.3$),
L:($u^*= 6.5 $, $\phi = 0.5$),
M:($u^*= 8 $, $\phi = 0.5$),
N:($u^*= 10 $, $\phi = 0.5$),
O:($u^*= 15 $, $\phi = 0.5$).}
\label{pd}
\end{figure}

Figure \ref{pd} shows a sketch of the equilibrium phase diagram for the model potential described by Eq. \ref{potencialduro}, 
in the $u^*$--colloidal-packing-fraction ($\phi$) plane, calculated for $\kappa\sigma =6$ \cite{PRL_2006_96_018303,JCP_2006_125_094502}.
At low $\phi$ and $u^*$, the fluid is the most stable phase. 
At high $\phi$, there exist  several crystal structures. Two of them coexist with the fluid: a disordered face-centred-cubic solid (disordered fcc)
at low $u^*$ , and a CsCl-like structure at high $u^*$. In the disordered-fcc solid, particles arrange on 
an fcc lattice, but the sign of the charge each particle bears does not follow any order throughout the crystal. 
At high $u^*$ the fluid-CsCl coexistence opens up and the solid coexists with a very low density fluid (gas). Buried underneath the gas-solid coexistence, 
there is a metastable gas-fluid coexistence line, which is denoted in the phase diagram by a dashed line. 
In our simulation, we quenched a stable fluid to the state points
indicated by an asterisk and a letter in Fig.~\ref{pd},
and studied the structural evolution of the fluid in its way towards either a solid or a 
gel.

In order to observe a rare event such as crystal nucleation in a simulation of a modest number of particles it is 
necessary to use {\it ad hoc} simulation techniques. We have used two of them: Umbrella Sampling \cite{JCP_1992_96_04655} and Forward Flux Sampling \cite{PRL_2005_94_018104}. 
The Umbrella Sampling method (US) involves     
biasing the sampling of an equilibrium MC simulation in order to explore high free energy configurations 
(containing large pre-critical clusters).  We refer the reader to the review published in [10] 
for the method details.
The Forward Flux Sampling technique (FFS) uses an order parameter relevant for the transition to 
split the free energy landscape in regions divided by hyper-surfaces (or interfaces). 
In the FFS scheme, 
the calculation of the flux of formation of post-critical solid clusters in the liquid (number of post-critical clusters  
formed per unit of time and volume) is substituted by  
the flux of formation of small pre-critical solid clusters  times the probability
to form a post-critical cluster once one of these small pre-critical clusters has been formed.
In contrast to the US method, FFS can also be  used in strongly out-of-equilibrium situations, such as crystallization under shear \cite{JCP_2008_129_134704}. 
FFS was applied for the first time to crystal nucleation to calculate the homogeneous crystallization 
rate in molten NaCl \cite{JCP_2005_122_194501}. 

In order to study the liquid-to-solid transition path it is necessary to identify the particles belonging 
to a solid cluster. To do that, we use a local bond order parameter analysis \cite{JCP_1996_104_09932}. 
In our study we compute the $q_6$ complex vector for each particle, 
which is a function of the relative orientation of a particle with respect
to its neighbours (two particles are neighbours if they are closer than 1.33 $\sigma$, the first g(r) minimum).
The component $m$ of the vector associated with the $i^{th}$ particle is given by:
\begin{equation}
q^i_{l,m}={\frac{1}{N_i}} \frac{\sum^{N_i}_{j} Y_{l,m}({\bf r}_{ij})}{({\bf q}^i_l {\bf q*}^i_l)^{1/2}}, ~~~~~m = [-6,6]
\end{equation}
where, for our case, $l=6$, $N_i$ is the number of neighbors of particle $i$, 
$Y_{l,m}({\bf r}_{ij})$ is a
spherical harmonic function whose form depends on $l$ and $m$ and whose value depends on the relative orientation of
particles $i$ and $j$ (${\bf r}_{ij}$). In a perfect bcc or fcc crystal, all the particles have the same environment,
and, therefore, the scalar product between the vectors of any pair of particles is 1.
Two neighbour particles are considered to be ``connected'' if the 
scalar product of their normalised $q_6$ vectors exceeds a threshold of 0.7. 
If a particle has at least 9 connections it
is labelled as ``solid''. Any two solid-like particles closer than  1.3 $\sigma$ belong to the same cluster. 
In this way we identify all solid clusters present in the metastable fluid phase (the size of the  biggest one 
is the order parameter we have used to study crystal nucleation).  
It is important to note that our criterion to identify solid-like particles is blind to the type of solid lattice 
(bcc or fcc). This means that by biasing the system to grow solid clusters with our solid
criterion we are not forcing the appearance of any particular solid structure. 

To measure the degree of substitutional charge disorder within the growing solid 
clusters we use the following charge order parameter:

\begin{equation}
\xi = - \frac{1}{N} \sum^{N}_{i=1} \frac{1}{N_{i}} \sum^{N_{i}}_{j=1} \alpha_i \alpha_j
\end{equation}

where $N$ is the number of particles in the solid cluster, $N_{i}$ is the number of neighbours of particle $i$  
(for this purpose, a neighbour is any particle at a centre-of-mass distance closer 
than 1.12$\sigma$), and $\alpha$ is either +1 or -1 depending on whether the particle is
positively or negatively charged. Thus, $\xi$ is equal to 1 if all particles are surrounded by oppositely charged
neighbours and -1 if all particles are surrounded by equally charged neighbors. 
Hence, for a perfectly ordered phase one should expect $\xi$ = 1, whereas for a random phase
$\xi$ should be 0. Of course, when thermal fluctuations are present, $\xi$ is
lower than 1 even if the solid phase is ordered. However, on average, $\xi$ will always be higher for a substitutionally ordered phase than for a disordered one. 
(Note that the threshold used to identify neighbouring particles here (1.12 $\sigma$) is different from that 
used for the orientational order parameter described above (1.33 $\sigma$). 
Both thresholds have been set to maximize the difference in order parametre value
between either "liquid" and "solid" particles or substitutionally ordered and
substitutionally disordered clusters).

When studying crystallization at low and constant densities ($NVT$ ensemble) we have used a looser criterion to detect solid-like 
particles. The reason is that in those conditions there are many particles at
the interface between a dense phase  and a gas phase.  
Obviously, an interface particle can not have as many solid ``connections'' as a bulk particle does.
For us, a particle is solid-like if it is ``connected'' to at least 50\% 
of its neighbours. Two particles are ``connected'' if the scalar product of their $q_6$ vectors is larger than 0.5. 
The neighbouring distances are defined in the same way as they were for the liquid-to-solid transition study at high densities (see above).
We use the $NVT$ ensemble for the study of the system's evolution at low densities because we are interested in the formation of gel-like structures. These structures are formed in (gas-solid) coexistence regions of the phase diagram. Since in a$p-T$ projection of the phase diagram there are not coexistence regions --but coexistence lines-- the $NpT$ensemble is not suitable if one desires to keep the system at coexistence.

\subsection{Experiments}

We used two different batches of polymethylmethacrylate (PMMA) particles, which were covalently labelled with the fluorescent dyes rhodamine isothiocyanate (RITC) and 7-nitrobenzo-2-oxa-1,3-diazol (NBD), and sterically stabilized with poly-12-hydroxystearic acid. We synthesized these particles by means of dispersion polymerization \cite{JCIS_2002_245_292}. Both particle types were $2.5 \mu m$  in diameter, with a size polydispersity of 3 \%, as determined from static light scattering measurements.

We suspended the particles in a mixture of as received cyclohexylbromide (CHB, Fluka) and 26.5\% cis-decalin (Sigma-Aldrich) 
by weight. This mixture nearly matched the refractive index and density of the particles and had a dielectric constant of 5.6. 
We made suspensions with a 1:1 number ratio of the differently labeled particles 
at an overall volume fraction of $\phi \approx 0.20$, 
and either with ($0.5 \mu M$) or without the charge determining tetrabutylammonium bromide 
salt (TBAB, Sigma-Aldrich) \cite{N_2005_437_7056,JPCM_2003_15_S3581}. 
In CHB, PMMA particles acquire a positive charge. When TBAB salt is added a fraction 
of the ions adsorbs onto the particle, with more of the small Br$^-$ ions adsorbing 
than of the large organic TBA$^+$ ions. The initial positive charge and the charge 
sign reversal point differ for each particle batch. 
We then made gradient samples by filling a glass capillary 
(Vitrocom) partially with the salt-free suspension and partially with the $0.5 \mu M$ TBAB suspension, 
and allowing it to form a macroscopic salt gradient over a couple days' time.
Suspending the particles in a salt gradient is a 
convenient way to quickly explore different experimental conditions.
We studied these samples by means of confocal microscopy.

\section{Results}
\subsection{Homogeneous crystal nucleation at high temperature and constant pressure}

The most widespread view of crystal nucleation is that provided by Classical Nucleation
Theory (CNT) \cite{kelton}. According to CNT, the formation of a crystal nucleus in a metastable liquid costs free energy due to 
the formation of an interfacial area $A$. On the other hand, 
the fact that the solid's ($sol$) chemical potential ($\mu$) is lower than that of the metastable fluid ($flu$),
is the driving force for crystal nucleation, 
and lowers the free-energy of crystal formation.
The balance between these two terms gives rise to a free energy barrier ($\Delta G$):

\begin{equation}
\Delta G = A \gamma - n |\mu_{sol} - \mu_{mf}|
\end{equation}

where $\gamma$ is the average liquid-solid surface free energy and $n$ is the number of particles contained in the largest solid cluster.
Because of the nucleation barrier, a (metastable) fluid can exist at state points of the phase diagram where the solid has a lower chemical potential.
Given that a metastable fluid is a quasi-equilibrium state, equilibrium thermodynamic statistics can be applied to study its properties. 
This allowed Duijneveldt and Frenkel to develop the US method to calculate nucleation free energy barriers \cite{JCP_1992_96_4655}. 
The US method relies on the assumption that pre-critical clusters are in quasi-equilibrium with 
the surrounding liquid. 
Hence, it is possible to use any kind of MC sampling scheme to analyse the cluster's properties (structure, shape, critical cluster size...)
via an US scheme. Here, we use two types of MC
simulations to study the properties of the pre-critical clusters: (i) MC $NPT$ simulations, in which volume moves and single particle displacements
are used to sample the configurational space, and (ii) MC $NPT$ simulations with charge swap moves, in which, besides the volume and particle moves, 
we occasionally try  (20\% of the times) to swap the positions of two oppositely charged particles. 
As mentioned before, as long as the quasi-equilibrium assumption holds, the properties of the clusters should not be affected by 
the way the system is sampled. 

\begin{figure}
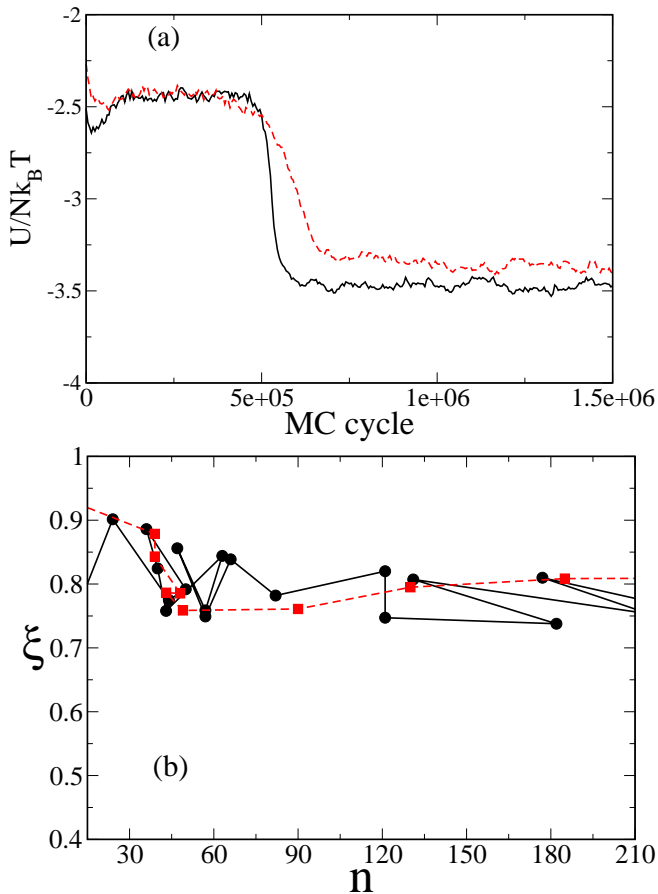
 
\includegraphics[clip,width=0.48\textwidth,angle=-0]{UvsMCcycleT05.eps}
\includegraphics[clip,width=0.48\textwidth,angle=-0]{espontaneaT05.eps}
\caption{\small 
(a) Internal energy $U$ versus the number of MC cycles, and (b) charge order parameter $\xi$ versus the number of particles in 
the largest solid cluster, n, for two liquid-to-solid trajectories, where the line connects two consecutive points along the trajectory. All results are obtained from two MC NPT simulations at state point A, where  
the solid line (circles) correspond to sampling without charge swap moves, while the dashed lines 
(squares) denote the ones with swap moves.}
\label{aT05}
\end{figure}

The US sampling scheme is tailored for conditions at which there is a high free energy barrier (few tens of $k_BTs$) separating 
the metastable fluid and the solid.
If the free energy barrier is not too high --nor too small (otherwise there would not exist a metastable liquid)-- one can observe the liquid-to-solid 
transition without any biased simulation technique. This is precisely the case of state point A, which is analysed in Fig.\ref{aT05}.
Fig.\ref{aT05} (a) shows the evolution of the internal energy per particle for two MC simulations (with and without charge swap moves) starting 
from a state-point-A metastable fluid configuration. After an initial stage in which the energy fluctuates around an 
average value (metastable liquid), the liquid transforms into a solid --as the energy drop proves. 
The solid formed in both cases (with and without charge swaps) corresponds to the stable phase at the state point A: CsCl.
Let us now analyse the nucleation pathway in more detail. In particular, we focus on the size and charge order of the 
largest cluster. 
Figure \ref{aT05} (b) shows the charge order parameter as a function of the size of the largest 
solid cluster along the transition path 
for both types of simulations.
No significant difference is found between the two cases. Hence, the way the system is sampled does not affect the nucleation path in state A, confirming 
the hypothesis that clusters are at any time in quasi-equilibrium with the surrounding metastable liquid. 
Nothing unexpected has been observed for the study of state point A. 

\begin{figure}
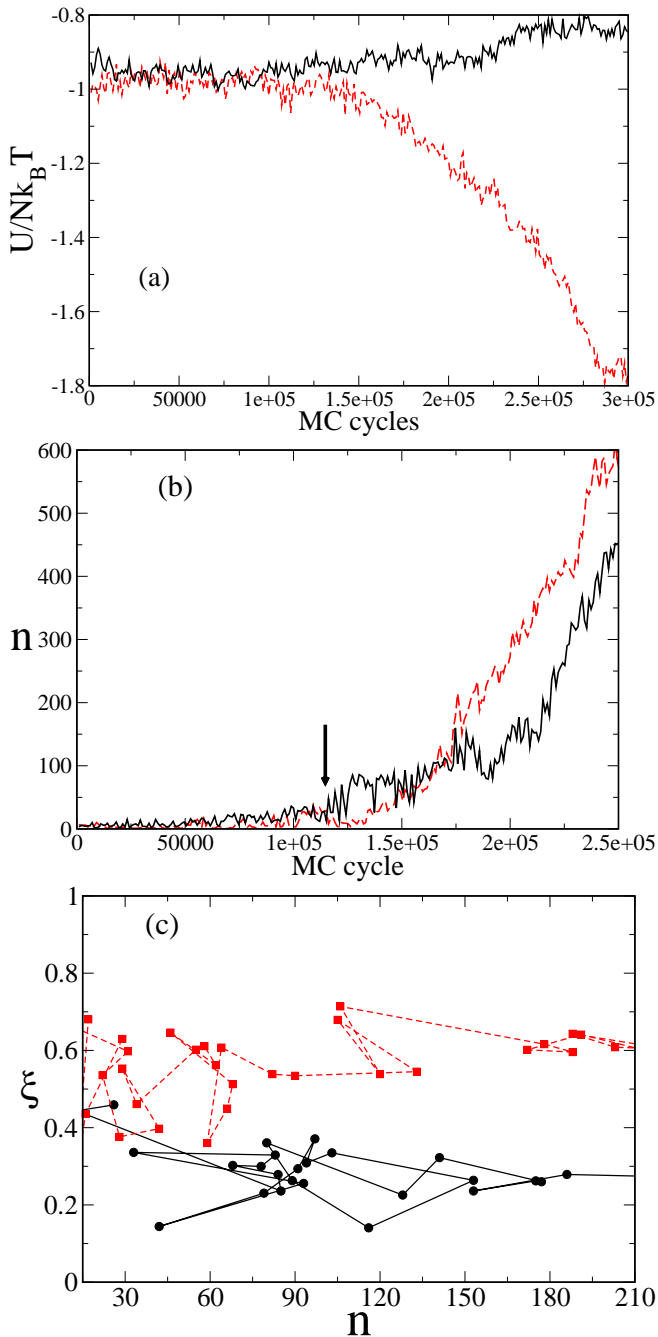
 
\includegraphics[clip,width=0.48\textwidth,angle=-0]{energy_vs_Nsteps_T1.eps}
\includegraphics[clip,width=0.48\textwidth,angle=-0]{N_vs_Nsteps-con-sin-cambio.2.eps}
\includegraphics[clip,width=0.48\textwidth,angle=-0]{conswap_T1.eps}
\caption{\small 
(a) Internal energy $U$ versus number of MC cycles, (b) Number of particles in the largest cluster, n, versus 
number of MC cycles, where the arrow indicates the point beyond which the liquid-to-solid 
trajectory has been analysed, and (c) Charge order parameter $\xi$ versus number of particles
in the largest solid cluster, n, for two liquid-to-solid trajectories, where the line connects consecutive points
along the trajectory. All results are obtained from two simulations at state point B, where the 
solid line (circles) correspond to sampling without charge swap moves, while the dashed lines (squares)
denote the ones with swap moves.}
\label{aT1}
\end{figure}

Now we focus our attention on state point B (also at a supersaturation high enough as to allow nucleation to occur 
spontaneously in the course of a regular $NpT$ simulation). 
In contrast to state point A, there is a competition between CsCl and disordered fcc 
because of the proximity of state B to the solid-solid coexistence line. 
Figure \ref{aT1} shows a different scenario as compared to the case
of state A. Depending on the MC simulation type, the metastable fluid ends up forming either an ordered CsCl phase 
(with charge swap moves) 
or a disordered-fcc lattice (without charge swap moves).
Note that the energy of the disordered-fcc solid (formed in the absence of swap moves)
is higher than that of the liquid, indicating  the entropic character of the
phase transition. The number of particles in the largest solid cluster, $n$,
as a function of the number of MC cycles is shown in Fig. \ref{aT1} (b). We have analysed the solid clusters 
when $n$ starts to grow continuously in size (indicated by an arrow in Fig.\ref{aT1} (b)).
Figure \ref{aT1} (c), representing the charge order parameter as a function of the 
cluster size along the nucleation pathway, 
shows that not only a different phase is obtained by changing the sampling, but also that the nucleation 
path is different: if no charge swap moves are included in the MC simulations solid clusters have a noticeably lower charge order parameter. 
This is an indication that a kinetic effect is playing an important role on the crystallization path. If the transition path was solely determined by
the underlying free energy landscape, pre-critical clusters would have had the same charge order parameter regardless of the sampling scheme. 

\begin{figure} \includegraphics[clip,width=0.48\textwidth,angle=-0]{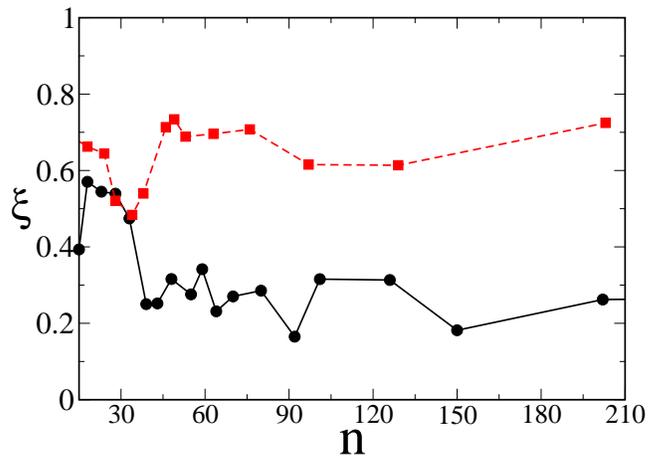}
\caption{\small  Charge order parameter $\xi$ versus number of particles n in the largest solid cluster for two 
liquid-to-solid trajectories at state point C (both calculated by means of the FFS method\cite{PRL_2005_94_018104,PRL_2007_99_055501}) with (squares) and without (circles) charge swap moves. The line connects consecutive points along the trajectory.}
\label{orderFFS}
\end{figure}

Some of the authors have recently reported a study of the liquid-to-solid transition at state point C \cite{PRL_2007_99_055501}, which is in line 
with the observations 
described here 
for state B. Since the supersaturation in state point C is lower than in B, a rare event simulation technique was required to observe the liquid-to-solid transition in 
computer simulations. Using the FFS technique we observed that the structure of the clusters along the transition path depends on the sampling scheme. 
If charge swap moves are included, the radial distribution function of the 
particles in the cluster reveals a CsCl-like structure, whereas an fcc lattice is observed otherwise \cite{PRL_2007_99_055501}. 
Figure \ref{orderFFS} shows the charge order parameter versus the number of particles in the largest solid cluster for two typical transition paths obtained in the FFS 
calculations. 
Confirming the study at state point B, the path obtained with charge-swap moves contains clusters with higher charge order. 
For small clusters the value of $\xi$ is similar.
Therefore, the difference between the two paths is not determined by the initial fluctuations towards the solid, but rather 
by the way particles are incorporated into the growing cluster. 
In order to grow an ordered cluster, particles have to be incorporated with a given charge sign at a given
location at the cluster surface before the cluster itself re-dissolves. This condition is only fulfilled if swap-moves
are included.

\begin{figure} \includegraphics[clip,width=0.48\textwidth,angle=-0]{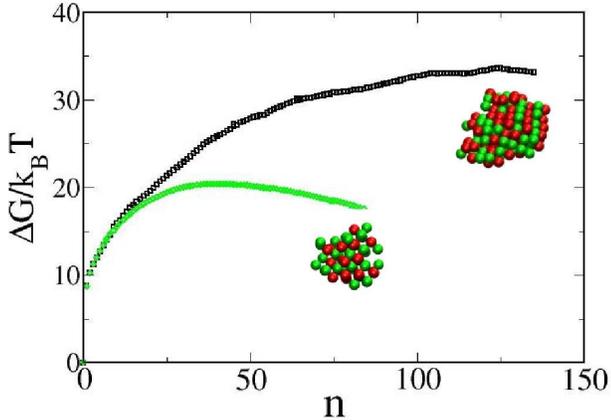}
\caption{\small Gibbs free energy $\Delta G$ as a function of cluster size n for CsCl ordered clusters (lower barrier)
and for disordered-fcc clusters (higher barrier) \cite{PRL_2007_99_055501}. 
A typical configuration of each type of cluster (disordered-fcc or ordered-CsCl) is shown in the 
figure. Green and red particles are oppositely charged.}
\label{barreras}
\end{figure}

In a recent paper, we have calculated the free energy barrier associated with each type of cluster (ordered CsCl/disordered-fcc) via US simulations \cite{PRL_2007_99_055501}. The barriers are shown in Fig. \ref{barreras}. 
We can grow either ordered or disordered clusters by respectively including or leaving out charge 
swaps moves in the US MC simulation.
This allows us to calculate the free energy associated with each path. 
As Fig. \ref{barreras} shows, a path containing disordered-fcc clusters has a higher free energy.
Since the US scheme is based on equilibrium simulations, 
the minimum free energy path should be eventually reached even if the system is inefficiently sampled 
(not including charge swapping).
In fact, for a window centred around 30 particles (the size beyond which the free energy associated with each 
type of path diverges),
around $3\cdot10^5$ MC cycles are required to complete a transition from disordered to ordered clusters.
This ``time'' is much larger than the time it takes for the clusters to grow and follow either type
of path (see Fig.\ref{aT1} (b) and (c)).
Therefore, without charge-swap moves, the clusters keeps growing with charge disorder with no time 
to equilibrate to the lowest free energy path.  

In summary, the simulations show that
both for states B and C the transition path depends on the mobility of the particles.
A possible interpretation is that, without the charge swap moves, 
the fluid is not ergodic on the time scale of cluster growth, 
not allowing for a minimization of the free energy of the path. 
The Stranski-Totomanow \cite{Ostwaldrevisited} conjecture, 
which states that the critical cluster structure is dictated by the 
lowest free energy barrier, is not compatible with our observations. For the system we present here, the 
mobility of the particles, and therefore the ability to equilibrate,  not only determines which phase is eventually formed, 
but also the transition path.

Since the no-charge-swap sampling is obviously more realistic in terms of dynamics, we expect that if an experiment were to be carried out under the same 
thermodynamics conditions, 
the system would follow the high free energy  route 
of disordered-fcc clusters. 
We have confirmed by means of BD simulations --which provide a more realistic description of the dynamics 
than MC simulations--
that a liquid quenched to state point B transforms into a disordered-fcc solid.

\subsection{Crystallization vs. vitrification at low temperature and constant volume}

\begin{figure} \includegraphics[clip,width=0.48\textwidth,angle=-0]{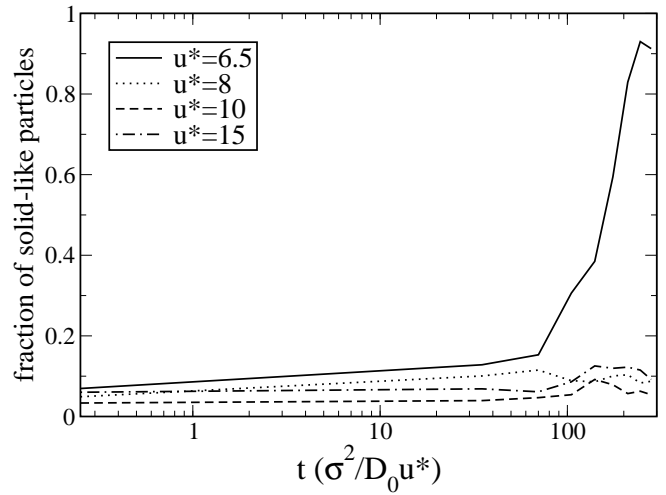}
\caption{\small Fraction of solid-like particles as a function of time along the course of a BD $NVT$ for states L-O (see Fig.\ref{pd}).  The starting configuration is a homogeneous fluid.}
\label{pack05cristal}
\end{figure}

So far we have presented results on the fluid-to-solid transition at 
high temperature and constant 
pressure.
Let us now analyse how the fluid evolves when we quench the system to high $u^*$ (low
temperatures) at constant volume (state points D-O).
At the temperatures corresponding to state points A-C the solid coexists with a 
high density fluid. By contrast, it coexists with a low density gas at the temperatures
corresponding to states D-O. 
According to the equilibrium phase diagram, a BD $NVT$ simulation starting from a homogeneous fluid
quenched to states D-O 
should show how the system phase separates into a low density gas and a CsCl crystal. 
However, at the end of our simulations ($\sim 300 t^*$) only 
state L became crystalline. In Fig. \ref{pack05cristal} the fraction of solid-like particles (see ``Methods and technical details'' section for the
criterion to define a particle as ``solid-like'') is plotted against time for the quenches at  packing fraction 0.5 (states L-O).
It is interesting to note that crystallization is only observed at state point (L).
This is somewhat counter-intuitive, since one expects 
faster crystallization for deeper quenches (higher thermodynamic driving force for crystallization).
Hence, it must be the slowing down of 
particles' dynamics that prevents the system from crystallizing at states M-O.
We do not claim that crystallization will never take place at states M-O --it will eventually (as a diamond will eventually become
graphite). We simply point out that the absence of crystallization within our simulation time for states M-O  
is not because of a high nucleation barrier, but because particles can not move enough to rearrange into a crystalline lattice.
Experiments \cite{S_2002_296_104}, theory \cite{PRL_2002_88_098301,L_2003_19_4493,S_2002_296_104}, and simulations 
\cite{PRL_2002_88_098301,S_2002_296_104,JPCM_2004_16_S4849} have confirmed the existence of an attractive glass for 
short-ranged attractive particles in a similar region of the phase diagram.

\begin{figure} \includegraphics[clip,width=0.48\textwidth,angle=-0]{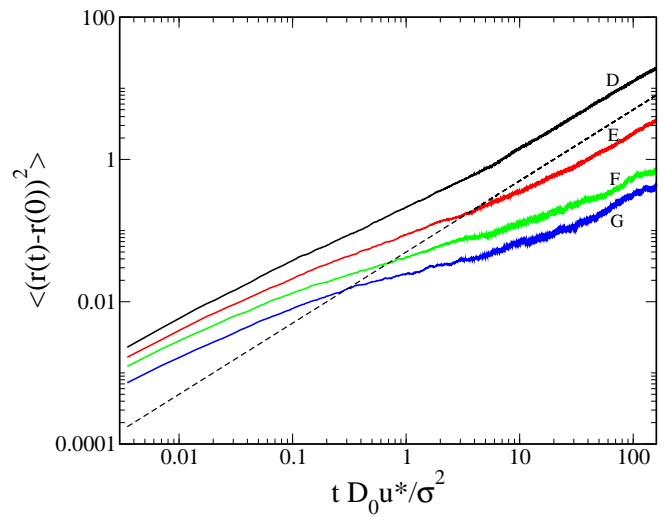}
\caption{\small Mean square displacement $\langle (r(t)-r(0))^2 \rangle$ for the structures formed 950 t* after the quench to state points D-G (see Fig.\ref{pd}). The dashed line --just for visual reference-- has slope 1.}
\label{msd}
\end{figure}

At lower packing fractions ($\phi =$ 0.1 and 0.3, states D-K) there is no evidence of crystallization even after $\sim 1000 t^*$. Fig. \ref{msd} shows that  the mean square displacement calculated for the structures obtained
950 $t^*$ after the quench develops a clear inflection point when increasing $u^*$, which is a footprint of sub-diffusive dynamics. 
The referred slowing down of particles' dynamics could be responsible for the absence of crystallization for quenches as deep as 
those to states G, K or O.

\begin{figure} 
\includegraphics[clip,width=0.48\textwidth,angle=-0]{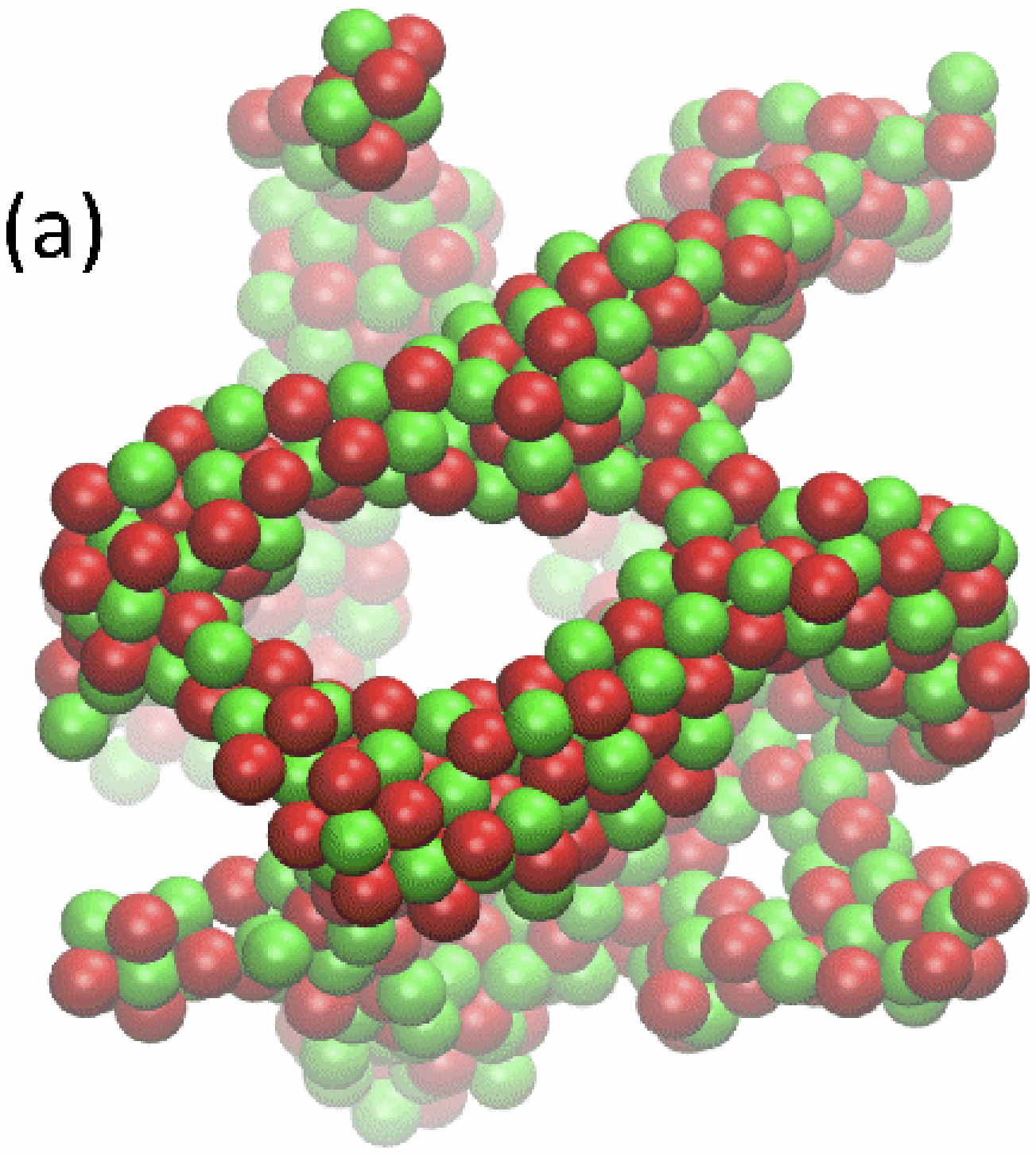}
\includegraphics[clip,width=0.48\textwidth,angle=-0]{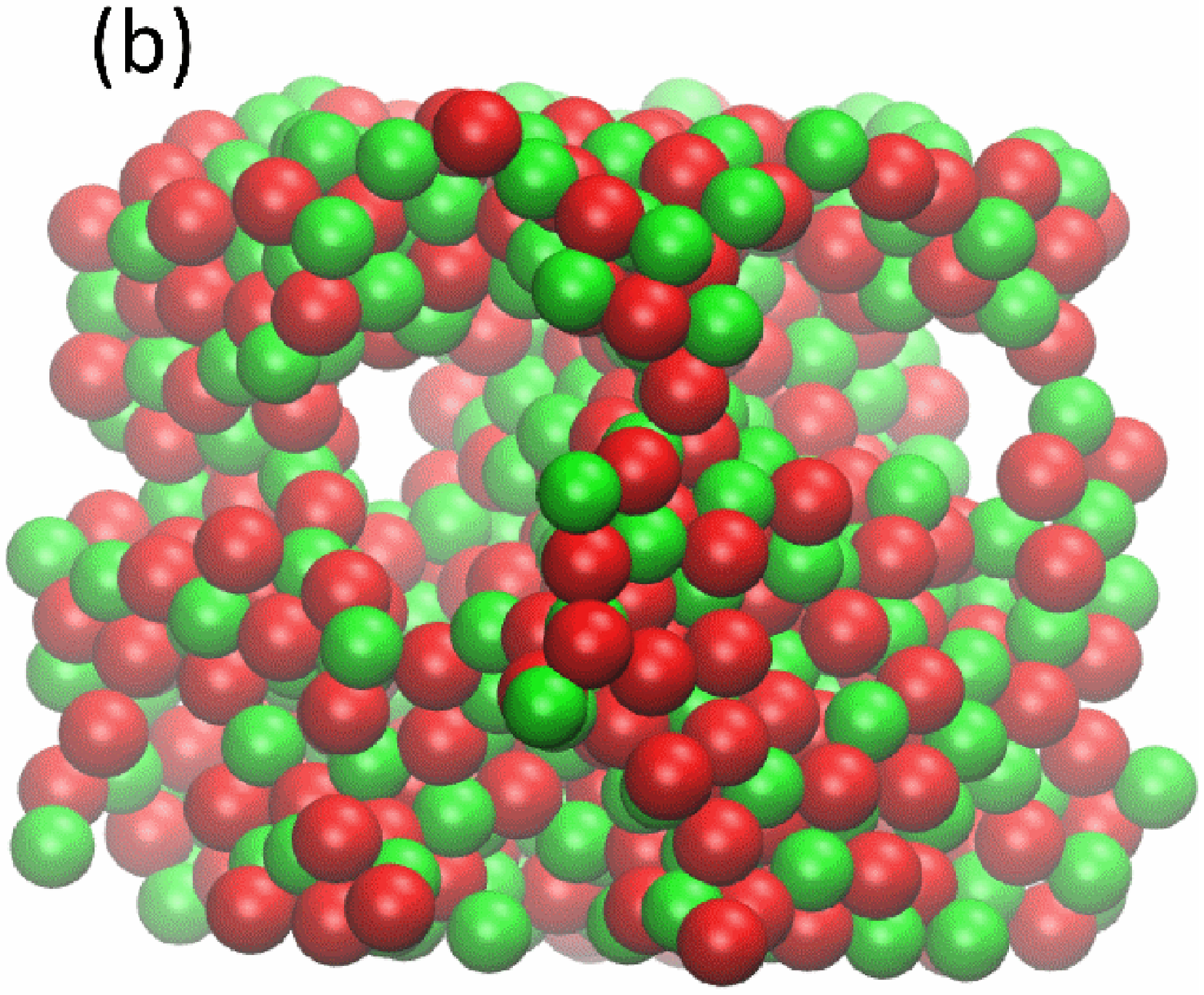}
\caption{\small Snapshot of the structures obtained 500 t* after the quench to states G (a) and K (b). 
Green and red particles are oppositely charged.}
\label{gelfoto}
\end{figure}

Fig. \ref{gelfoto} shows snapshots of the system at state points G and K $500 t^*$ after the quench. We remind the reader that the initial configuration
of our simulations is a homogeneous fluid phase suddenly quenched to states G and K.
Both configurations can be described as percolating networks of high density amorphous branches, 
which corresponds to the typical description of a gel. 
For a percolating network of dense amorphous branches to be a gel, it has to possess a non zero yield stress. 
In a recent paper some of the authors have shown experimentally that oppositely charged colloids can indeed form gels \cite{tobegelcharged}. 
In order to do so we imaged, by means of a confocal microscopy, a colloidal network 
while imposing a solvent flow through the sample. 
The network resisting the drag force exerted by the solvent demonstrates the ability of 
oppositely charged particles to form gels. 
In the same 
study, we demonstrated that the morphology of the gel-like configurations (Fig. \ref{gelfoto})  
is due to an interruption of the metastable gas-liquid coexistence region~\cite{tobegelcharged}.   
In the equilibrium phase diagram, Fig. \ref{pd}, there is a gas-liquid metastable coexistence 
buried underneath the gas-solid coexistence region.
When a homogeneous fluid is quenched to states D-K a gas-liquid spinodal decomposition starts immediately.
As soon as particles gather in high density regions, the spinodal coarsening process slows down 
because the local dynamics of the particles gets slow 
(due, in turn, to the high energy at contact and  the short interaction range).  
The restricted mobility of the particles not only 
slows down the spinodal coarsening (giving rise to gel-like networks), but also prevents crystallization.

\begin{figure} \includegraphics[clip,width=0.48\textwidth,angle=-0]{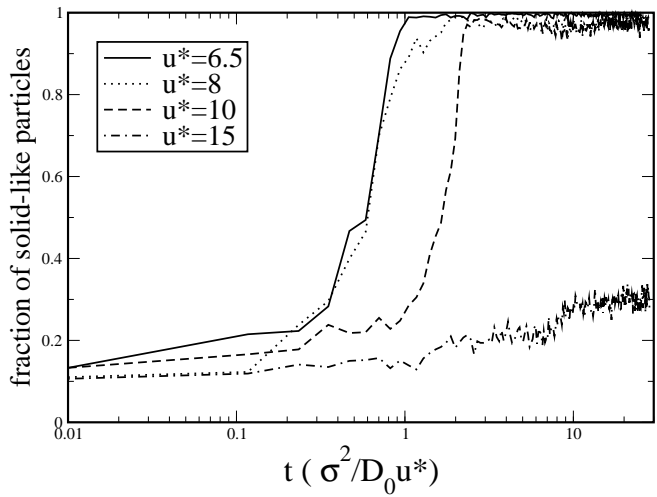}
\caption{\small Same as Fig.\ref{pack05cristal} but with a longer interaction range: $\kappa \sigma = 2$ instead of 6.}
\label{nvstimeks2}
\end{figure}

Finally, we analyse the influence of the range of the potential on the ability to crystallize. 
This information can be helpful for experimentalists, for making good colloidal crystals depends sensitively on numerous experimental variables~\cite{N_2005_437_7056,PRL_2005_95_128302}.
In order to assess the role of the interaction range on crystallization we have performed BD $NVT$ simulations
at the same packing fraction and $u^*$ as states L-O but imposing a longer screening length ($\kappa \sigma = 2$ instead of 6).
Starting from a homogeneous configuration we monitor the fraction of crystalline particles over time (see Fig. \ref{nvstimeks2}).
Whereas for $\kappa \sigma = 6$ we only observed crystallization at $u^* = 6.5$ (Fig. \ref{pack05cristal}), now the system readily 
becomes solid  for  $u^* = 6.5$, $u^* = 8$ and $u^* = 10$. 
Therefore, for the same contact energy, increasing the interaction range favours crystallization. 
This can be intuitively understood: particles in a dense amorphous phase need to rearrange in order to 
form crystalline domains; given that the attraction force is higher the shorter the interaction range (for a given $u^*$), 
rearrangements are easier for longer ranged interactions.
From an experimental point of view
increasing the interaction range in a system of charged particles implies decreasing the amount of salt in the medium. 
Fig. \ref{nvstimeks2} also shows a manifestation of glassy particle dynamics (in the same way as Fig. \ref{pack05cristal} does 
for $\kappa \sigma = 6$). Again,  on a thermodynamic basis, one would expect that increasing the contact energy
favours the formation of the solid phase. 
On the contrary, the system becomes glassy for high contact energies and crystallization gets hindered. 
For the quench at $u^* = 15$, a crystal cluster of $\sim 250$ particles forms at the initial stage of the simulation.  
The fact that this crystalline domain can not propagate throughout the simulation box is an indication that the system 
has fallen out of equilibrium.

\begin{figure} 
\includegraphics[clip,width=0.48\textwidth,angle=-0]{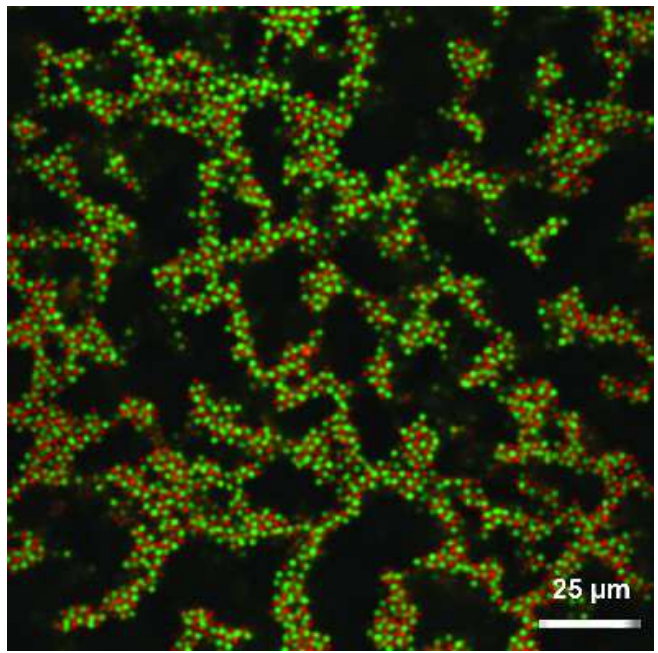}
\caption{\small A confocal microscopy image of a gel-like structure with amorphous branches readily formed after
vigorously shaking the sample with a suspension of the two particle species. Green and red particles are oppositely charged.}
\label{expperamorphous}
\end{figure}

\begin{figure} 
\includegraphics[clip,width=0.48\textwidth,angle=-0]{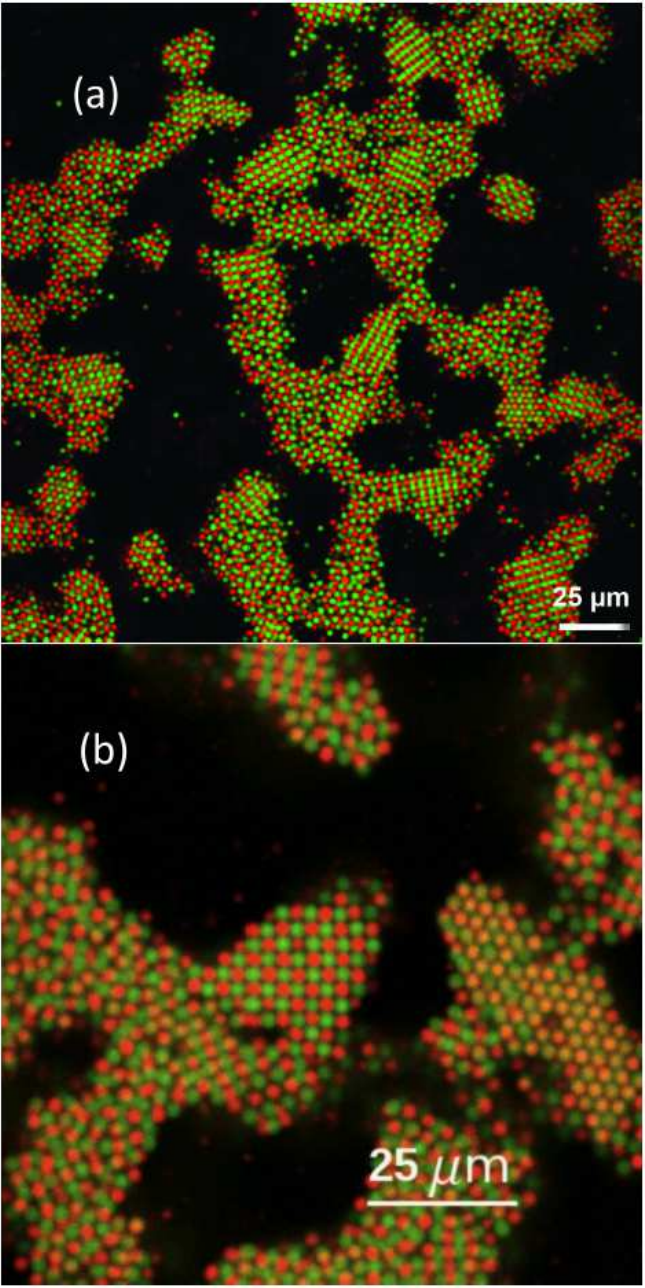}
\caption{\small (a) A confocal microscopy image of a gel-like structure with crystalline branches observed two days after mixing. 
(b) A zoomed-in confocal image. Crystalline regions are observed in contact with amorphous ones.}
\label{expperccristal}
\end{figure}

As shown above, at low $\phi$ the system forms a percolating network of dense 
amorphous branches when quenched to a  metastable gas-liquid coexistence point. 
In view of our observations for $\kappa \sigma =2$ it is reasonable to expect crystallization 
within such amorphous branches at a larger screening lengths. In that case, 
a percolating network of crystalline branches would result. Such a structure  
has already been observed for attractive particles \cite{PRE_2001_64_021407,PRE_2007_75_011507,PRE_2007_76_041402}. 
Some of us have recently  studied the case of $\kappa \sigma = 2$ at packing fraction $0.1$
for contact energies ranging from $u^*=12.5$ to $u^*=30$ \cite{tobegelcharged}. At $u^*=20$ we found that about $15\%$ of the 
system was crystalline; for the rest of the investigated states the system remained amorphous by the end of our 
simulations. 
Nevertheless, we show in our experiments (see experimental details in the ``Methods and technical details'' section) that it is possible to 
obtain a percolating network of crystalline branches for oppositely charged particles (Fig.\ref{expperccristal}). 
In the experiments, we observed a percolating network of amorphous branches (Fig. \ref{expperamorphous})
minutes after putting in contact both low and high-salt concentration suspensions. 
These gels were observed in the high-salt-concentration half of the sample (particles are not
oppositely charged in a free-of-salt environment \cite{N_2005_437_7056}).  
Two days later, when the salt gradient had smoothened, 
we observed that the branches had coarsened and (partially) cristallized (Fig. \ref{expperccristal} (a) and \ref{expperccristal} (b)). 
Hence, the formation of percolating crystalline structures 
is due to local crystallization in the course of a metastable spinodal decomposition (by analogy with an amorphous-branches' gel, 
whose formation can be described as vitrification in the course of a metastable spinodal decomposition).
Further experimental work needs to be done to more accurately determine how far the time scale of gelation and crystallization were actually apart, 
as this was not investigated yet. 
In contrast to the experiments shown here, in a recent work,  
we did not observe crystallization in the branches of oppositely-charged-particles' gels \cite{tobegelcharged}. 
An important difference between the two experimental systems was the amount of salt  present in the medium.  
In the current experiments there is initially a salt gradient along the sample ($0.47 \mu M - 0 \mu M$). 
Hence, the average salt concentration was $\sim 0.23 \mu M$, whereas it was $1 \mu M$ for the experiments reported in \cite{tobegelcharged}.
The particles' size was $2.5 \mu m$ for the present experiments ($pr$) and $2 \mu m$ for the former ones ($for$). 
Using the Derjaguin-Landau-Verwey-Overbeek (DLVO) \cite{APURSS_1941_14_633,BookVO} approximation for $\kappa$ ($\kappa=\sqrt{8\pi \lambda_B \rho_{salt}}$) we 
can estimate the ratio between both $\kappa \sigma$: $\kappa \sigma_{for}/\kappa \sigma _{pr} \approx 1.7$.
Therefore, the interaction range was larger for the present experiments. This is in agreement with our simulations, where we 
observe that crystallization is easier for $\kappa \sigma = 2$ than for $\kappa \sigma = 6$. Summarizing, both simulations and experiments suggest that
colloidal crystals form more readily for systems with long screening lengths.
However, more careful experiments need to be carried out, 
since in our experimental system the charges of the colloidal particles
are coupled to the concentration of salt in the medium.

\section{Conclusions}

We present a study of the evolution of a fluid of oppositely charged particles after being quenched to a state point of
the equilibrium phase diagram where either a solid or a solid coexisting with a gas is the stable phase.
  
First of all, we have studied, by means of Monte Carlo simulations at constant temperature and pressure,
crystal nucleation for quenches to state points where the solid is the most stable phase. 
Interestingly, the crystallization path taken by a metastable liquid close to an
ordered-solid/disordered-solid coexistence line depends on the simulations' sampling scheme.
If a ``fast'' sampling scheme is used (including charge-swap moves), the nucleation path is a succession of
ordered-solid clusters, while if a ``slow'' sampling is performed (not including charge-swap moves), 
the nucleation path
contains disordered-solid clusters. By means of Umbrella Sampling calculations we have measured the free energy
associated with each type of path: The free energy of the ordered-clusters path is lower than that of the disordered clusters.
We interpret our results to indicate a lack of ergodicity of the slowly sampled fluid on the time scale
of crystal growth. Our results contradict the Stranski-Totomanow conjecture, which states that the transition
path is determined by the minimum free-energy barrier for nucleation. For the case presented here,
the mobility of the particles
plays a major role in the selection of the nucleation path.
We argue that if an experiment were to be carried out under the same thermodynamic conditions, the system would
follow the higher free energy route of disordered clusters.
If the same study is repeated at a state point far away from the ordered-solid/disordered-solid coexistence line,
the sampling scheme does not affect the transition path. 

Secondly, we have studied the evolution of a fluid quenched to state points of the equilibrium phase diagram where a gas and  a solid coexist.
In order to do so, we have carried out Brownian Dynamics simulations at constant volume and temperature.
We observe clear signs of slow dynamics. For instance, the system does not crystallize faster for deeper quenches --which is what
would be expected if crystallization is controlled by thermodynamics (nucleation free-energy barrier).
Moreover, the mean square displacement reveals sub-diffusive dynamics:
a clear inflection point develops with the quench depth.
The morphology of the system is very much influenced by the presence of a metastable gas-liquid spinodal:
Percolating networks of dense glassy branches (gels) are obtained as a consequence of the arrest of an ongoing spinodal
decomposition. Finally, we have studied the influence of the interaction range on crystallization.
Increasing the interaction range, while keeping the contact energy and the density constant,
favors crystallization over vitrification. We believe this is the case because particles can rearrange
more easily the longer the interaction range is. Nevertheless, the slow-dynamics footprint of
crystallization being delayed by an increased quench depth, is also present when the interaction range is increased.
Crystallization in the experiments might be due, in agreement with our 
computer simulations, to an increase of screening length as the salt concentration decreases. 
However, this has to be further investigated, since the amount of charge 
on the particles is also correlated with the salt concentration in the system.
Our work illustrates that either amorphous or crystalline-branched networks can be seen as an interrupted ongoing spinodal decomposition. In the former case, 
the interruption is due to vitrification, whereas in the latter it is due to crystallization. 

\section{Acknowledgements}
This work was financially supported by the Nederlandse
Organisatie voor Wetenschappelijk Onderzoek (NWO) and by the Stichting voor Fundamenteel Onderzoek der Materie (FOM).
We  acknowledge D. Derks for the synthesis of RITC labeled particles.
Finally, some of us gratefully acknowledge the Spanish national team  for beating Germany in the 2008 Eurocup final, which was a source of
inspiration and a stimulus to finish this work; however, some of the other authors couldn't care less.

\end{document}